\documentclass[twocolumn,prl,amsmath,amssymb,superscriptaddress]{revtex4-1}

\usepackage{graphicx}
\usepackage{dcolumn}
\usepackage{bm}
\usepackage{soul}
\usepackage{color}
\usepackage{epstopdf}
\usepackage[version=3]{mhchem}
\usepackage{lipsum}
\usepackage[outercaption]{sidecap}
\usepackage{floatrow}
\begin{document}

\title{Theory of pressure-induced rejuvenation and strain-hardening in metallic glasses}

\author{Anh D. Phan}
\affiliation{Faculty of Materials Science and Engineering, Computer Science, Artificial Intelligence Laboratory, Phenikaa Institute for Advanced Study, Phenikaa University, Hanoi 12116, Vietnam}
\email{anh.phanduc@phenikaa-uni.edu.vn}
\affiliation{Department of Nanotechnology for Sustainable Energy, School of Science and Technology, Kwansei Gakuin University, Sanda, Hyogo 669-1337, Japan}
\author{Alessio Zaccone}
\affiliation{Department of Physics "A. Pontremoli", University of Milan, via Celoria 16, 20133 Milano, Italy}
\affiliation{Cavendish Laboratory, University of Cambridge, JJ Thomson Avenue, CB3 0HE Cambridge, United Kingdom}
\affiliation{Statistical Physics Group, Department of Chemical Engineering and Biotechnology, University of Cambridge, Philippa Fawcett Drive, CB3 0AS Cambridge, United Kingdom}
\email{alessio.zaccone@unimi.it}
\author{Vu D. Lam}
\affiliation{Institute of Materials Science, Vietnam Academy of Science and Technology, Hanoi, Vietnam}
\affiliation{Graduate University of Science and Technology, Vietnam Academy of Science and Technology, Hanoi, Vietnam}
\author{Katsunori Wakabayashi}
\affiliation{Department of Nanotechnology for Sustainable Energy, School of Science and Technology, Kwansei Gakuin University, Sanda, Hyogo 669-1337, Japan}
\date{\today}

\date{\today}

\begin{abstract}
We theoretically investigate high-pressure effects on the atomic dynamics of metallic glasses. The theory predicts compression-induced rejuvenation and the resulting strain hardening that have been recently observed in metallic glasses. Structural relaxation under pressure is mainly governed by local cage dynamics. The external pressure restricts the dynamical constraints and slows down the atomic mobility. In addition, the compression induces a rejuvenated metastable state (local minimum) at a higher energy in the free energy landscape. Thus, compressed metallic glasses can rejuvenate and the corresponding relaxation is reversible. This behavior leads to strain hardening in mechanical deformation experiments. Theoretical predictions agree well with experiments.
\end{abstract}

\maketitle

Metallic glasses have become promising materials for high-performance engineering due to their fascinating mechanical properties \cite{13,22,23,24}. Metallic glasses are produced by very rapid solidification of metal alloys to avoid the crystalline state, and therefore they are electrically conducting materials with internal atomic structure which is fully disordered like in a glass. To understand the properties of amorphous materials for several applications, it is essential to understand the underlying relaxation dynamics in connection with the free energy landscape. 

The structural ($\alpha$) relaxation time, $\tau_\alpha$, depends on temperature, external pressure, and confinement effects \cite{13,63,64,31}. At high temperatures, the atomic mobility is high and materials are in a liquid-like state. On the other hand, a significant slowing down of dynamics upon cooling, reflected in a steep increase of $\tau_\alpha$, leads to solid materials with extraordinary mechanical properties in terms of both hardness and ductility. In particular, thermodynamic and physical properties including viscosity and diffusivity \cite{63,31,39} change  abruptly by orders of magnitude around the glass transition temperature $T_g$.
The glass state obtained in this way occupies a deep metastable minimum in the energy landscape (EL), often called 'inherent state'. Upon mechanical deformation, the yielding of metallic glass is often ductile with a shear-softening stress-strain curve which goes through a yielding maximum before reaching a Newtonian-like viscous plateau~\cite{Johnson,Schall,Falk},  qualitatively similar to the response of hard-sphere colloids~\cite{Egelhaaf}. This behavior can be understood in terms of the strain driving the system out of the inherent state through an energy barrier (thus reversing the $\alpha$ process towards a more fluid state) followed by a downhill trajectory in the EL~\cite{Lacks1,Lacks2}.

Recent experiments from different groups have shown that applying an external pressure to metallic glasses leads to some spectacular and unexpected physics, for which a mechanistic understanding is missing. Shock compression of metallic glass leads to significant glass rejuvenation, which is signalled by substantial increase in the excess relaxation enthalpy and in the decrease of the boson peak in the specific heat (where the boson peak represents an excess of low-energy vibrational states over the Debye law in the vibrational density of states)~\cite{37}. Even more recently, Greer and co-workers~\cite{30} have demonstrated experimentally that compression-induced rejuvenation goes along with strain-hardening in the mechanical deformation. This implies that, upon increasing the deformation, the material does not fully yield to plastic flow and the stress (instead of dropping with the increase of strain and levelling off in the viscous plateau) keeps increasing indefinitely with the increase of strain. 

In this letter, we present a self-contained theoretical description of the EL of metallic glasses under compression. The theory is able to predict the emergence of a higher energy metastable state in the EL as a consequence of the applied pressure, which explains the rejuvenation phenomenon. Furthermore, the predicted pressure-induced EL is able to explain the strain-hardening effect observed experimentally, and for which no theoretical explanation was at hand.

We theoretically investigate the activated processes and structural rearrangements in metallic glasses under compression using the Elastically Collective Nonlinear Langevin Equation (ECNLE) theory \cite{2,6,7,8,10,42,11,40,41,34,35}. All substances are described as a dense hard-sphere fluid with  particle diameter, $d$, and number of particles per volume, $\rho$. At ambient pressure, mobility of a tagged particle is mainly affected by the nearest-neighbors interactions. This dynamics at a fixed temperature $T$ is quantified via the dynamic free energy, which is given by~\cite{Viehman}
\begin{eqnarray}
\frac{F_{dyn}(r)}{k_BT} &=& \int_0^{\infty} dq\frac{ q^2d^3 \left[S(q)-1\right]^2}{12\pi\Phi\left[1+S(q)\right]}\exp\left[-\frac{q^2r^2(S(q)+1)}{6S(q)}\right]
\nonumber\\ &-&3\ln\frac{r}{d}, 
\label{eq:2}
\end{eqnarray}
where $\Phi = \rho\pi d^3/6$ is the volume fraction, $k_B$ is Boltzmann constant, $r$ is the displacement of the tagged particle from its initial position, $q$ is the wavevector, $S(q)$ is the static structure factor calculated using the Percus-Yevick theory \cite{1}. 

The first term on the r.h.s. of Eq. (\ref{eq:2}) describes how the tagged particle interacts with neighbors in the first shell (caged dynamics) and favors localized states. Thus, this is strongly dependent on the structural input. While the second term corresponds to a dilution entropy which tends to delocalize the particles.

When the density of the material is low, $F_{dyn}(r)$ decreases with increasing $r$ and the particles are not dynamically arrested \cite{3,4}. At higher density, the configurational freedom is lowered due to reduction of free-volume. A free-energy barrier for the hard-sphere fluid emerges at $\Phi \geq 0.432$ and the onset of the slow dynamics is observed \cite{3}. The tagged particle is dynamically swept within a cage formed by its neighbors, which probes the EL of the system. The dynamic free energy profile provides key physical quantities: the localization/caging length, $r_L$, corresponding to the first minimum of $F_{dyn}(r)$, and the barrier position, $r_B$, located at the maximum of $F_{dyn}(r)$. From these, one can calculate a jump distance and a primary barrier height by $\Delta r =r_B-r_L$ and

$F_B=F_{dyn}(r_B)-F_{dyn}(r_L)$, respectively.

A further contribution to the caging may come from long-ranged elastic correlations~\cite{8,11,41,35}, however in the present analysis of experimental data we found this additional contribution to be unnecessary. Detailed discussions can be found in the Supplemental Material \cite{43}. Based on Kramers' theory~\cite{Conchuir}, the structural ($\alpha$) relaxation time is
\begin{eqnarray}
\frac{\tau_\alpha}{\tau_s} = 1+ \frac{2\pi}{\sqrt{K_0K_B}}\frac{k_BT}{d^2}\exp\left(\frac{F_B}{k_BT} \right),
\label{eq:6}
\end{eqnarray}
where $K_B$=$\left|\partial^2 F_{dyn}(r)/\partial r^2\right|_{r=r_B}$ is absolute curvature at $r_B$ and $\tau_s$ is a short atomic-vibration time scale. The explicit expression of $\tau_s$ for various thermal liquids, polymers, and amorphous drugs is given elsewhere \cite{6,7,11,41,42,35}. In this work, we continue to use this value for calculations of metallic glasses.

To determine the temperature dependence of our ECNLE structural relaxation time, we also employ thermal-expansion analysis to map from a hard-sphere density to temperature. The thermal mapping is~\cite{11,42,41,35}
\begin{eqnarray}
T \approx T_0 - \frac{\Phi - \Phi_0}{\beta\Phi_0}.
\label{eq:7}
\end{eqnarray} 
where $\beta \approx 12\times 10^{-4}$ $K^{-1}$ is a common volume thermal expansion coefficient for many organic materials and $\Phi_0 \approx 0.5$ is a characteristic packing fraction. Although the $\beta$ value is higher than the thermal expansion coefficient of metallic glasses, there are two main reasons for us to keep using this for the thermal mapping: (i) Our ultimate goal is to propose a universal mapping having a fixed set of parameters to investigate various types of amorphous materials; (ii) precise agreement with real experimental values should not be expected because hard-sphere ECNLE calculations do not take into account interatomic interactions. Thus, all information of these interactions and their consequences is encoded in the $\beta$ and $T_0$ parameter.

\begin{figure}[htp]
\center
\includegraphics[width=9cm]{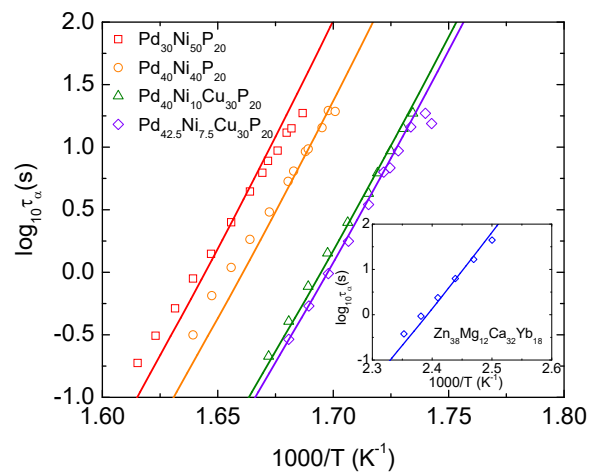}
\caption{\label{fig:2}(Color online) Logarithm of structural relaxation time of several metallic glasses as a function of $1000/T$ at ambient pressure. Open points are experimental data in Ref. \cite{9,12} and solid curves correspond to ECNLE calculations.}
\end{figure}

Figure \ref{fig:2} shows the theoretical and experimental temperature dependence of $\tau_\alpha$ for several metallic glasses at atmospheric conditions. The ECNLE theory predicts the glass transition temperature (defined by $\tau_\alpha(T_g) = 100$s) $T_g = 588.5$, 582.37, 570.31, 569.3, and 398.176 $K$ for \ce{Pd_{30}Ni_{50}P_{20}}, \ce{Pd_{40}Ni_{40}P_{20}}, \ce{Pd_{40}Ni_{10}Cu_{30}P_{20}}, \ce{Pd_{42.5}Ni_{7.5}Cu_{30}P_{20}}, and \ce{Zn_{38}Mg_{12}Ca_{32}Yb_{18}}, respectively, while their experimental counterparts are 586, 582, 570, 569, and 395 $K$ \cite{9,12}. We find $\tau(\Phi_g \approx 0.6585)=100$s and  $T_0\approx T_g+(\Phi_g-\Phi_0)/\beta\Phi_0$ can be directly determined using the experimental $T_g$. Thus, our numerical results agree quantitatively with experiments without any adjustable parameter. This finding suggests that the activated events are mainly governed by the nearest-neighbor shell interactions.

To capture pressure effects on the glass transition of metallic glasses, we suppose that an external pressure, $P$, acts on a tagged particle and generates mechanical work when undergoing a small displacement in the compression field. The localized free volume of the particle is $\Delta V(r) \approx rd^2$. By assuming that the pressure remains uniform throughout the material, this mechanical work is approximated by $P\Delta V(r) \approx Prd^2$ and modifies the effective non-equilibrium free energy as \cite{35,41}
\begin{eqnarray}
\frac{F_{dyn}(r)}{k_BT} = \frac{F_{dyn}(r,P=0)}{k_BT}+ \frac{P}{k_BT/d^3}\frac{r}{d}.
\label{eq:newFdyn}
\end{eqnarray}

\begin{figure}[htp]
\center
\includegraphics[width=9cm]{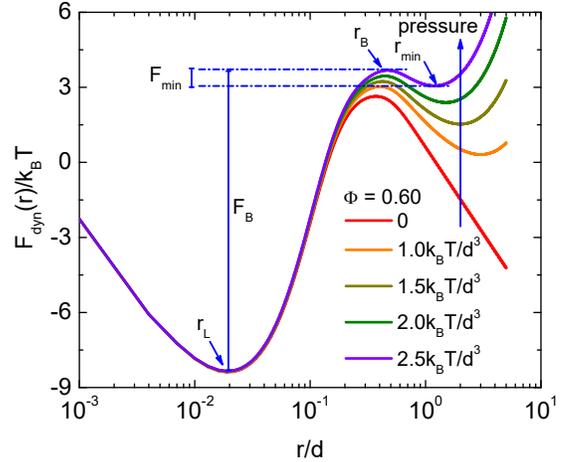}
\caption{\label{fig:1}(Color online) The dynamic free energy as a function of reduced particle displacement for a hard sphere fluid of packing fraction $\Phi = 0.60$ at several pressures in unit of $k_BT/d^3$. Key length scales barriers for the local dynamics are indicated.} 
\end{figure}

Figure \ref{fig:1} shows the dynamic free energy for a hard-sphere fluid of $\Phi=0.60$ at several pressures. At ambient pressure (very small pressure), one can approximate $P \approx 0$ and observe a single energy minimum in Fig. \ref{fig:1}. This implies the local hopping is an irreversible process. When external compression is applied, the localization length/barrier position barely changes. Effects of the nearest-neighbor constraints on the energy landscape are significantly weakened as the particle diffuses away from its cage, while the role of applied pressure becomes greater. Pressure-induced hindrance of the diffusion causes another localized state. Consequently, a second local minimum position, $r_{min}$, appears. By ignoring the dynamic caging constraint in $F_{dyn}(r)$, one analytically finds $r_{min} = \frac{3k_BT/d^2}{P}$. The analysis is quantitatively consistent with full numerical results. Clearly, $r_{min}$ is density-independent and is shortened with increasing compression, while an increase of external pressure reduces the secondary barrier height, $F_{min}=F_{dyn}(r_B)-F_{dyn}(r_{min})$. At extremely high compression, the second localized state disappears and the tagged particle is dynamically arrested for a very long (practically infinite) time. An increase of $\Phi$ raises $F_B$ but leaves the qualitative behaviour of the free energy unaltered \cite{43}.


This dynamic energy landscape behaves as a two-level system and it is possible for the escaped particle to return to its initial configuration. The double well leads to well-tempered metadynamics and reversibility of structural relaxation. The emergence of a higher-energy secondary minimum qualitatively explains compression-induced rejuvenation \cite{37,28} and reversible structural relaxation \cite{28,29,20} in metallic glasses. An increase of pressure promotes the glass rejuvenation by generating the higher-energy secondary minimum towards which the glass can rejuvenate. This analysis can qualitatively explain recent experimental results in Ref. \cite{37}. Authors in Ref. \cite{37} also found that the atomic structures undergoing rejuvenation are disordered in a complicated manner associated with a nanometers length scale. This idea seems plausible since the ECNLE theory predicts that $r_{min}$ is several times greater than the particle diameter $d$. Based on previous studies \cite{8,10,2,35,41}, $d\approx0.4-1.2$ nm for various amorphous materials.

The ECNLE theory agrees with the fact that many-body atomic rearrangements cause the irreversible structural relaxation \cite{23,24,25,26,27}. By using the hard-sphere model, we neglect all chemical and conformational complexities, but still achieve the two-level system as shown in Fig. \ref{fig:1}. Thus, influence of chemical ordering on the reversible relaxation phenomenon due to pressure is relatively minor. These conclusions are consistent with the findings of Refs. \cite{27,25}.

\begin{figure}[htp]
\center
\includegraphics[width=9cm]{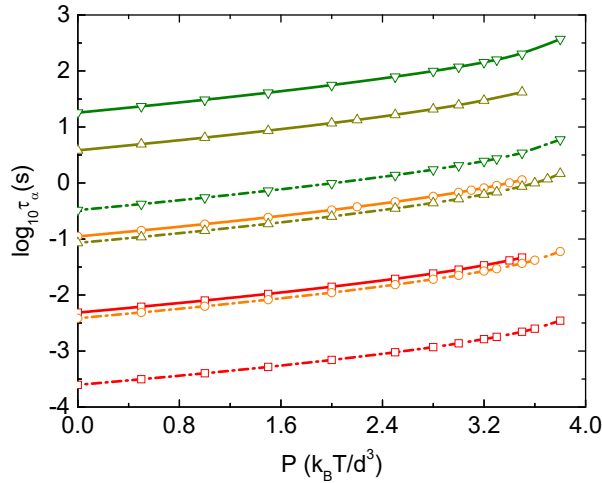}
\caption{\label{fig:3}(Color online) The theoretical pressure dependence of structural relaxation time of \ce{Pd_{30}Ni_{50}P_{20}} (solid curves) and \ce{Pd_{40}Ni_{10}Cu_{30}P_{20}} (dashed-dotted curves). Red, orange, yellow-green, and green curves corresponds to calculations at $T =$ 635.33, 618.67, 602, and 595.33 $K$ respectively. }
\end{figure}

Figure \ref{fig:3} shows the response of structural relaxation process in \ce{Pd_{30}Ni_{50}P_{20}} and \ce{Pd_{40}Ni_{10}Cu_{30}P_{20}} to the external pressure at isothermal condition. Here, we still use Eq. (\ref{eq:7}) to convert from a packing fraction of an effective hard-sphere fluid in ECNLE calculations into temperature, and suppose that parameters of this thermal mapping are independent of pressure \cite{41,35}. These assumptions have been used in previous works \cite{41,35} and successfully described $\tau_\alpha(T,P)$ of amorphous drugs \cite{41}, polymers \cite{35}, and a simulated organic glass-forming liquid \cite{21}. The external compression induces more restriction on motions of a single particle within its cage and slows down the structural mobility, leading to $\alpha$-relaxation time increasing with pressure consistent with prior observations~\cite{Wei-Hua}. From these ECNLE results, one can also find pressure-induced strong-to-fragile transitions in glass-forming liquids \cite{35,21}, which was experimentally reported in Ref. \cite{18,19}. 

In a different interpretation, compression induces both a free volume reduction and the collective long-range structural rearrangement of atoms \cite{16,37}. At lower pressures, atoms become more localized and the effects of the free volume annihilation dominate \cite{16}. This assumption is consistent with ECNLE predictions since $r_L$ decreases with compressing and the secondary minimum of $F_{dyn}$ barely occurs. Thus, the relaxation enthalpy of the system is decreased \cite{16,17}. However, at higher pressures, external compression densifies the atomic structure by reorganizing atoms into new configurations \cite{16}, and the dynamics is slowed down. A further increase of atomic packing density forces the interatomic distance to be shorter, thus enhancing the interaction energy and the relaxation enthalpy. Now, metallic glasses can rejuvenate to higher energy states, consistent with experimental observations in Refs. \cite{16, 37}. 
This result provides a theoretical explanation to the spectacular increase of relaxation enthalpy upon compression measured experimentally upon shock-compression of metallic glasses~\cite{37}.

\begin{figure}[htp]
\center
\includegraphics[width=9cm]{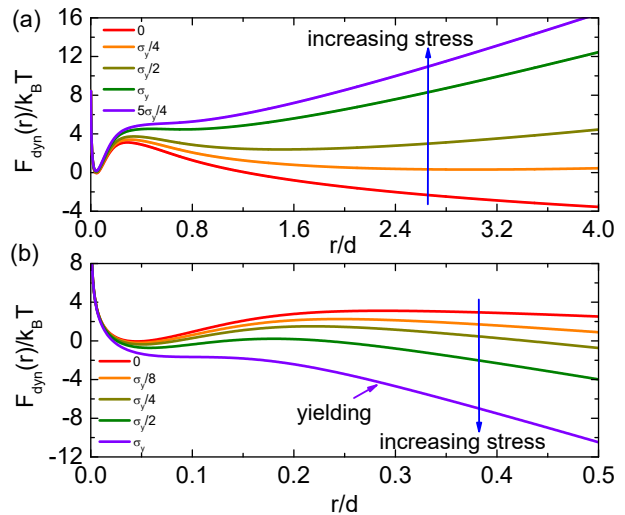}
\caption{\label{fig:4}(Color online) The dynamic free energy as a function of reduced particle displacement for a hard sphere fluid of packing fraction $\Phi = 0.53$ at several stresses in the case of (a) strain-hardening and (b) strain-softening. The yield stresses, $\sigma_y$, of strain-hardening ($\sim$ 26 $k_BT/d^3$) and strain-softening ($\sim$ 4 $k_BT/d^3$) calculations correspond to the minimum stress to  suppress the secondary and primary localization in $F_{dyn}(r)$, respectively.}
\end{figure}

This ECNLE approach can also be used to understand stress-strain relations in the mechanical response of glasses. Figure \ref{fig:4}a shows the emergence of a precompression-induced secondary minimum in the free energy landscape of the strained glass, calculated using Eq.(\ref{eq:newFdyn}). Crucially, the vanishing of the secondary minimum as $\sigma \geq \sigma_y\approx 26k_BT/d^3$ leads to a more stable localized state.  This finding clearly explains why rejuvenated metallic glasses due to pre-compression exhibit the strain-hardening behavior during mechanical tests such as those in Ref. \cite{30}.

In contrast, when metallic glasses are stretched without an applied pressure or pre-compression (hence without the compression-induced rejuventation), Eq.(\ref{eq:newFdyn}) can be modified as $F_{dyn}(r,\sigma) = F_{dyn}(r,\sigma=0)-\sigma d^2r$. Figure \ref{fig:4}b shows how the dynamic free energy evolves by increasing applied stress. In this case, the standard strain-softening behavior of atomic and molecular glasses is retrieved~\cite{Lacks1,Lacks2}. For $\sigma \geq \sigma_y\approx 4k_BT/d^3$, at the yielding point, the primary localization minimum vanishes and $\tau_\alpha = \tau_s$ is constant and independent of the tension. This coincides with the onset of plastic flow or steady state with increasing strain in the post-yield regime.  

Tensile deformation of rejuvenated metallic glasses in Ref.\cite{30} reveals the same strain hardening as observed in compressive testing. This behavior can be explained via the energy profile $F_{dyn}(r,\sigma) = F_{dyn}(r,\sigma=0)+P^*d^2r-\sigma d^2r$, here $P^*$ is a pre-compression pressure. Since the pre-compression of the metallic glasses in Ref. \cite{30}, before machining, is greater than spatial elongation during tension, it seems that $P^*\geq \sigma$ and the rejuvenated state is maintained during the subsequent tensile test. Overall, we have shown that the strain-hardening behavior originates from the pressure-induced rejuvenation that modifies the underlying dynamic free energy landscape. 

In conclusion, we have presented a dynamical theory for pressured-induced rejuvenation and strain-hardening in metallic glasses. At ambient pressure, the irreversible $\alpha$ activated event is caused by local microstructural rearrangements. The presence of compression raises the local barrier height and slows down the glassy dynamics of metallic glasses. When the applied pressure is large enough, the emergence of a higher-energy metastable state is predicted, thus leading to rejuvenation. The theory can also predict key features of stress-strain relations in the mechanical deformation of metallic glasses at high pressure. In the absence of compression, the predicted free-energy landscape is consistent with strain-softening and yielding, as typically observed with metallic glasses~\cite{Johnson}. Strikingly, in the presence of compression-induced rejuvenation, the theory predicts a free energy landscape for deformation which is consistent with strain-hardening, whereby a transitory softening is followed by hardening upon deforming the material further. This prediction provides a theoretical explanation to the recently discovered strain-hardening in metallic glasses due to compression-induced rejuvenation in~Ref. \cite{30}. All in all, the presented framework may lead to a new mechanistic understanding of amorphous materials at high pressure by establishing a powerful connection between the underlying energy landscape and the resulting macroscopic properties.

\begin{acknowledgments}
This work was supported by JSPS KAKENHI Grant Numbers JP19F18322 and JP18H01154. This research was funded by the Vietnam National Foundation for Science and Technology Development (NAFOSTED) under grant number 103.01-2019.318.
A.Z. acknowledges financial support from US Army Research Office through contract nr. W911NF-19-2-0055.
\end{acknowledgments}

\end{document}